\documentclass[pdflatex,iicol,sn-mathphys-num]{sn-jnl}

\usepackage{multirow}
\usepackage{amsmath,amssymb,amsfonts}
\usepackage{amsthm}
\usepackage{mathrsfs}
\usepackage[title]{appendix}
\usepackage{textcomp}
\usepackage{manyfoot}
\usepackage{booktabs}
\usepackage{algorithm}
\usepackage{algorithmicx}
\usepackage{algpseudocode}
\usepackage{listings}
\usepackage{soul}
\usepackage[normalem]{ulem}
\usepackage{graphicx}
\usepackage{dcolumn}
\usepackage{bm}
\usepackage[per-mode=symbol]{siunitx}
\DeclareSIUnit{\bits}{bits}
\usepackage[T1]{fontenc}
\usepackage{float}
\usepackage{mathptmx}
\usepackage{makecell}
\usepackage{csquotes}
\usepackage{threeparttable}
\usepackage{nicefrac}
\usepackage{xcolor}
\usepackage{braket}
\usepackage{anyfontsize}
\usepackage{acronym}
\usepackage{lineno}

\renewcommand{\figurename}{\textbf{Fig.}}
\renewcommand{\tablename}{\textbf{Table}}

\begin{document}

\newacro{QKD}{quantum key distribution}
\newacro{QSCF}{quantum secure coin flipping}
\newacro{PM}{phase modulator}
\newacro{MZI}{Mach-Zehnder interferometer}
\newacro{QBER}{quantum bit error rate}
\newacro{FSSI}{free-space Sagnac interferometer}
\newacro{SPS}{single photon source}
\newacro{WCP}{weak coherent pulses}
\newacro{SKR}{secure key rate}
\newacro{QD}{quantum dot}
\newacro{AWG}{arbitrary waveform generator}
\newacro{SNSPD}{superconducting-nanowire single photon detector}
\newacro{NF}{notch filter}
\newacro{GTP}{Glan-Thompson polarizer}
\newacro{WPOL}{Wollaston polarizer}
\newacro{CBG}{circular Bragg grating}
\newacro{TDC}{time-to-digital converter}
\newacro{CW}{clockwise}
\newacro{CCW}{counterclockwise}
\newacro{IRF}{instrument response function}
\newacro{hBN}{hexagonal boron nitride}

\title{Ultrastable, low-error dynamic polarization encoding of deterministically generated single photons}

\author[1]{\fnm{Joscha} \sur{Hanel}}
\author[1]{\fnm{Zenghui} \sur{Jiang}}
\author[1]{\fnm{Jipeng} \sur{Wang}}
\author[1]{\fnm{Frederik} \sur{Benthin}}
\author[1]{\fnm{Tom} \sur{Fandrich}}
\author[1]{\fnm{Eddy Patrick} \sur{Rugeramigabo}}
\author[2]{\fnm{Raphael} \sur{Joos}}
\author[2]{\fnm{Michael} \sur{Jetter}}
\author[2]{\fnm{Simone Luca} \sur{Portalupi}}
\author*[1]{\fnm{Jingzhong} \sur{Yang}}\email{jingzhong.yang@fkp.uni-hannover.de}
\author[1]{\fnm{Michael} \sur{Zopf}}
\author[2]{\fnm{Peter} \sur{Michler}}
\author*[1,3]{\fnm{Fei} \sur{Ding}}\email{fei.ding@fkp.uni-hannover.de}

\affil[1]{\orgdiv{Institut für Festkörperphysik}, \orgname{Leibniz Universität Hannover}, \orgaddress{\street{Appelstr. 2}, \city{Hannover}, \postcode{30167}, \state{Niedersachsen}, \country{Germany}}}

\affil[2]{\orgdiv{Institut für Halbleiteroptik und Funktionelle Grenzflächen, Center for Integrated Quantum Science and Technology (IQ\textsuperscript{ST}) and SCoPE}, \orgname{Universität Stuttgart}, \orgaddress{\street{Allmandring 3}, \city{Stuttgart}, \postcode{70569}, \state{Baden-Württemberg}, \country{Germany}}}

\affil[3]{\orgdiv{Laboratorium für Nano- und Quantenengineering}, \orgname{Leibniz Universität Hannover}, \orgaddress{\street{Schneiderberg 39}, \city{Hannover}, \postcode{30167}, \state{Niedersachsen}, \country{Germany}}}


\abstract{
The ability to inscribe information on single photons at high speeds is a crucial requirement for quantum applications such as quantum communication and measurement-based photonic quantum computation. Nowadays, most experimental implementations employ \aclp{PM} in single-pass, \acl{MZI} or Michelson interferometer configurations to encode information on photonic qubits. However, these approaches are intrinsically sensitive to environmental influences, limiting the achievable quantum error rates in practice.\\
We report on the first demonstration of a polarization encoder for single-photon qubits based on a \acl{FSSI}, showcasing inherent phase stability and overcoming previous error rate limitations. Telecom-wavelength single photons emitted by a quantum dot are modulated by the encoder under a repetition rate of $152\,\unit{MHz}$. A \acl{QBER} of $0.69(2)\%$ is achieved, marking the lowest error rate reported to date for high-speed information encoding on single photons. This work represents a key advance towards robust, scalable, and low-error quantum information processing with single photon sources.
}


\maketitle

\section{Relevance of Photonic Qubits} \label{sec:Introduction}
The deterministic inscription of information onto individual qubits is a foundational capability for quantum information technologies. In particular, the ability to prepare photonic qubits in well-defined quantum states and to perform measurements in well-defined bases is essential for a broad range of applications. These include measurement-based quantum computing \cite{spedalieriHighfidelityLinearOptical2006, obrienOpticalQuantumComputing2007} and quantum-cryptographic schemes such as \ac{QKD} \cite{bennettQuantumCryptographyPublic2014, loMeasurementDeviceIndependentQuantumKey2012} and quantum bit commitment \cite{brassardQuantumBitCommitment1993, kentUnconditionallySecureBit2012}. Such schemes are expected to play a pivotal role in securing future digital infrastructures by enabling communication protected by information-theoretically unbreakable encryption. Beyond secure communication, these capabilities also open avenues towards other quantum application circumstances such as \ac{QSCF} \cite{bennettQuantumCryptographyPublic2014} and secure voting systems \cite{broadbentInformationTheoreticallySecureVoting2008}.\\
To date, \Ac{QKD} remains the most extensively tested and mature application of photonic quantum cryptography. It has been demonstrated across a wide range of protocols in terms of sources, either decoy-state \ac{QKD} based on \ac{WCP} \cite{sibsonIntegratedSiliconPhotonics2017, grunenfelderSimpleHighspeedPolarizationbased2018, avesaniResourceeffectiveQKDFieldtrial2021, avesaniFullDaylightQuantumkeydistribution2021, fan-yuanRobustAdaptableQuantum2022, liHighrateQuantumKey2023}, or true \acp{SPS}  \cite{intalluraQuantumKeyDistribution2007, intalluraQuantumCommunicationUsing2009, takemotoQuantumKeyDistribution2015, gaoQuantumKeyDistribution2022, samanerFreeSpaceQuantumKey2022, morrisonSingleemitterQuantumKey2023, al-jubooriQuantumKeyDistribution2023, zahidyQuantumKeyDistribution2024}. While decoy-state \ac{QKD} is currently ahead in terms of \ac{SKR} \cite{liHighrateQuantumKey2023} and are already commercially available, \acp{SPS} offers distinct advantages for the establishment of scalable quantum networks \cite{kimbleQuantumInternet2008, wehnerQuantumInternetVision2018, zopfEntanglementSwappingSemiconductorGenerated2019, bassobassetEntanglementSwappingPhotons2019, vanloockExtendingQuantumLinks2020}. For instance, protocols such as measurement-device-independent \ac{QKD} \cite{owenMDIQKDUsing2021} are expected to be carried out with \acp{SPS}, which can potentially offer performance advantages over \ac{WCP}-implementations \cite{zhouMeasurementdeviceindependentQuantumKey2018}. Recent theoretical and experimental studies have also shown that \acp{SPS} can surpass the bits-per-pulse transmission limit both in \ac{QKD} \cite{zhangExperimentalSinglePhotonQuantum2025c} and \ac{QSCF} \cite{vajnerSinglePhotonAdvantageQuantum2024}. The performance of \acp{SPS} has improved significantly in recent years \cite{bozzioEnhancingQuantumCryptography2022}, particularly in brightness \cite{nawrathBrightSourcePurcellEnhanced2023} and single-photon purity \cite{schweickertOndemandGenerationBackgroundfree2018, reindlHighlyIndistinguishableSingle2019, tommBrightFastSource2021}, allowing for long-distance \ac{QKD} in both laboratory settings \cite{takemotoQuantumKeyDistribution2015, morrisonSingleemitterQuantumKey2023} and real-world field trials \cite{zahidyQuantumKeyDistribution2024, yangHighrateIntercityQuantum2024}. \\
Nevertheless, several technical challenges remain for the practical use of \ac{SPS}, especially precise and low-loss modulation of single photons to encode information in phase, time-bin, or polarization degrees of freedom. Two critical constraints limit this capability: first, the need to minimize the loss of the modulation setup, demanding high-quality optical components; second, the requirement of constant electronic modulation signals imposed on the \ac{PM} during the decay lifetime of the used \ac{SPS} (usually on the order of hundreds of $\unit{ps}$), so that each emitted photon over this time receives identical modulation. These challenges are far less severe in \ac{WCP}-based systems, where setup losses can be mitigated with high-intensity sources and pulse durations can be easily reduced to tens of $\,\unit{ps}$ \cite{liHighrateQuantumKey2023}. Furthermore, for both \ac{SPS} and \ac{WCP} systems, long-term stability of the modulated quantum states is critical. To address this, various Sagnac interferometer-based encoders have been proposed \cite{wangPracticalGigahertzQuantum2018, liHighspeedRobustPolarization2019, agnesiAllfiberSelfcompensatingPolarization2019, avesaniStableLowerrorCalibrationfree2020} and implemented in \ac{WCP}-based \ac{QKD} experiments \cite{liuPolarizationmultiplexingbasedMeasurementdeviceindependentQuantum2018,  chenIntegratedSpacetogroundQuantum2021, maSimpleQuantumKey2021, avesaniResourceeffectiveQKDFieldtrial2021, fan-yuanRobustAdaptableQuantum2022, tangTimebinPhaseencodingQuantum2023}. Only very recently, this technique has been applied in an \ac{SPS}-based \ac{QKD} experiment for the first time \cite{xingjianPolarizationencodedQuantumKey2025}.\\
In this work, we present a \ac{FSSI} BB84 polarization encoder and demonstrate real-time polarization encoding of single photons from a \acf{QD} emitting at $1560.4(1)\,\unit{nm}$. A \ac{QBER} of $0.69(2)\%$ and an encoding agreement of $96.5(2)\%$ with theoretically expected polarization states are obtained from the polarization projection measurement. To our knowledge, this is the first demonstration of information encoding on single photons using an \ac{FSSI}, and the reported values represent the highest performance to date for dynamic polarization encoding with a \ac{SPS}. In this experiment, these results are achieved by using high-quality free-space polarization optics and a low $V_{\pi}$ electro-optical \ac{PM}, driven by well-defined electrical pulses from an \ac{AWG} operating at $152\,\unit{MHz}$. The total loss of the encoder module is measured to be 5.17(5)$\,\unit{dB}$. Finally, we demonstrate the long-term operational stability of this encoder by measuring the polarization state of a laser passing through the setup over a continuous period of 60 hours.

\section{Source excitation and characterization} \label{subsec:Excitation}
\begin{figure}
    \centering
    \includegraphics[width=\linewidth]{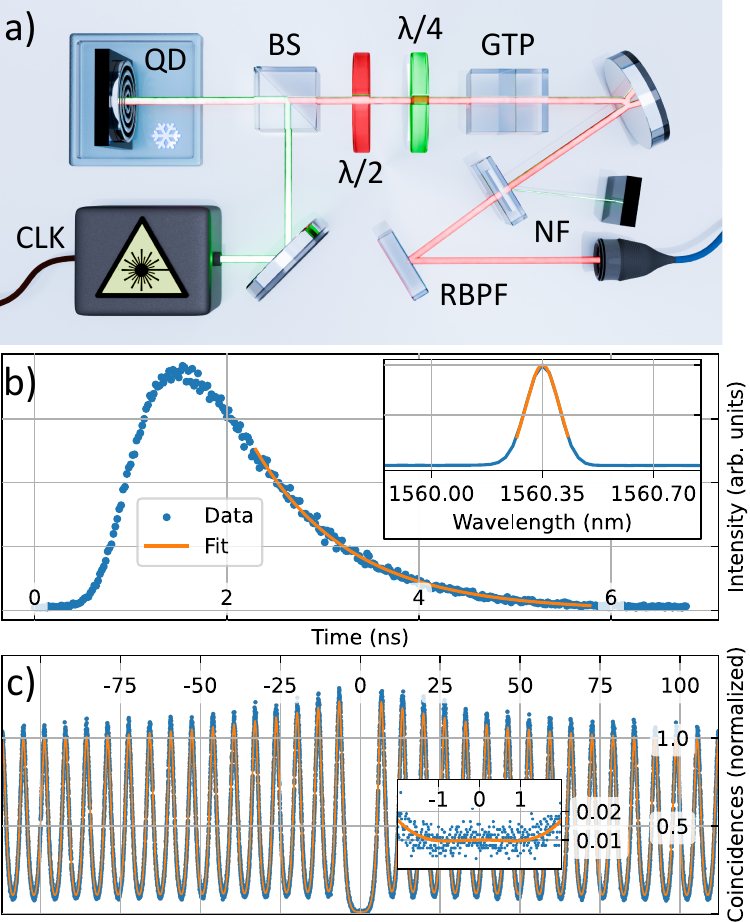}
    \caption{\textbf{Excitation, photoluminescence extraction and characterization of the \ac{QD}-\ac{SPS}.} a) Excitation and photoluminescence extraction setup. A $152\,\unit{MHz}$ modelocked laser is used for $p$-shell excitation of the \ac{QD}. The emitted photons are collected by an objective (not shown) with numerical aperture 0.7, polarized, spectrally filtered, and sent to the encoder module in a fixed polarization state. BS: 99/1 beamsplitter; GTP: Glan-Thompson polarizer; NF: notch filter; RBPF: reflecting bandpass filter; CLK: clock. b) Time-resolved fluorescence measurement of the \ac{QD} with an exponential decay fit on the falling edge, yielding a decay time of $\tau=990(7)\,\unit{ps}$. Inset: filtered, polarized emission spectrum of the QD (blue). By means of a Lorenzian fit (orange, ignoring the sides of the peak due to spectral cutoff by the RBPF), a central wavelength of $1560.4(1)\,\unit{nm}$ and a FWHM of $0.19(1)\,\unit{nm}$ are extracted. c) Fit (orange line) and data (blue dots) showing a second-order coherence measurement on photons emitted from the \ac{QD}. The fit yields a value of $g^{(2)}(0)=0.55(8)\%$ (see Methods section for details).}
    \label{fig:Characterization}
\end{figure}
In Fig. \ref{fig:Characterization}a), the excitation setup is depicted. Our \ac{SPS} consists of an InGaAs \ac{QD} embedded in a \ac{CBG} cavity \cite{nawrathBrightSourcePurcellEnhanced2023} and placed inside a cryostat (attoDRY 1100) at $3.8\,\unit{K}$. The \ac{QD} emits directly in the telecom C-band, making our setup suitable for use in data transmission over optical fiber. A modelocked laser (PriTel UOC) with a repetition rate of $152\,\unit{MHz}$, a pulse duration of $\sim10\,\unit{ps}$ and a wavelength of $1533.8(1)\,\unit{nm}$ is used for $p$-shell excitation of the \ac{QD}. Before being sent to the \ac{QD}, it passes through a tunable bandpass filter (not shown) to suppress laser background at the \ac{QD} emission wavelength. The emission from the \ac{QD} is spectrally filtered using three volume-Bragg-gratings (each with a nominal suppression of $>60\,\unit{dB}$). Two of these gratings act as notch filters supressing remaining laser light reflected by the sample (only one shown in Fig. \ref{fig:Characterization}a)), while the third reflects only the \ac{QD} emission to further eliminate remaining background at other wavelengths. Since the \ac{QD} emission is only partially polarized, the photons are additionally sent through a \ac{GTP} to obtain a fixed starting polarization state.\\
For characterization, the photons were then coupled to a single mode fiber and sent directly to a \ac{SNSPD} (Pixel Photonics PCU) or spectrometer (Princeton Instruments SpectraPro). A count rate of roughly $3.2\unit{Mcts/s}$ of spectrally filtered and polarized photons (including an estimated $85\%$ fiber coupling efficiency and a $37\%$ nominal \ac{SNSPD} detection efficiency) is achieved for an excitation power of $7\,\text{µW}$ (measured directly before the objective).
Autocorrelation (radiative decay data) are obtained by temporally correlating detection times between two \ac{SNSPD} channels in a Hanbury Brown and Twiss \cite{hanburybrownTestNewType1956} setup (between one channel and a reference signal from the \ac{AWG}) and shown in Fig. \ref{fig:Characterization} b) (c)).

\section{Encoding Scheme}
\subsection{Working Principle} \label{subsec:Encoding}
For all encoding measurements, the photons emitted from the \ac{QD} are routed through a fiber-based circulator to the \ac{FSSI} encoder module. Its fundamental working principle is similar to that of single-pass, \ac{MZI} or Michelson encoders: an input polarization state consisting of an equal superposition of $\ket{H}$ and $\ket{V}$ is prepared, and then a relative phase $\varphi$ between the two components is applied to generate the required output state. The key difference is that in the Sagnac configuration, both interferometer arms (i.e. propagation directions) consist of the same optical path, making the interferometer intrinsically phase stable. This is crucial when trying to obtain the desired output states, as unwanted relative phase variations would alter the output state and increase the \ac{QBER}. Note that both the \ac{CW} and \ac{CCW} components pass through the fibers and \ac{PM} in the same polarization, which is a requirement for the intrinsic phase stability as different crystal axes generally exhibit different temperature dependence of their refractive indices.\\
Here, four output polarization states from two bases are generated, satisfying the BB84 requirements \cite{bennettQuantumCryptographyPublic2014}. The input state is prepared by rotating the $\lambda/2$ waveplate (see Fig. \ref{fig:Encoder}a)) such that the two outputs of the Wollaston polarizer receive exactly equal power. This is calibrated using a continuous-wave laser (Toptica CTL-1550) tuned to the \ac{QD} emission wavelength. We label the input state the $\ket{D}$ state (with a relative phase $\varphi=0$), for convenience and without loss of generality. Accordingly, we label the other three generated states $\ket{R}$ ($\varphi=\pi/2$), $\ket{A}$ ($\varphi=\pi$) and $\ket{L}$ ($\varphi=-\pi/2$).\\
Each output of the Wollaston polarizer is then coupled to a polarization-maintaining fiber connected to the \ac{PM} (EOSPACE), and on each side a pair of waveplates (only one pair shown in Fig. \ref{fig:Encoder}) is used to align the polarization with the slow axis of the fiber.\\
Due to an additional $\Delta T\approx3.3\,\unit{ns}\approx\nicefrac{1}{2\cdot 152\,\unit{MHz}}$ delay in one of the arms, the early/\acf{CW} and late/\acf{CCW} photon components arrive at the \ac{PM} at different times, and their relative phase $\varphi$ can be modulated by changing the voltage applied to the \ac{PM} in between. For our \ac{PM}, voltages of $V_\pi=4.2\,\unit{V}$ and $V_{\pm\pi/2}=\pm2.1\,\unit{V}$ are applied in a repeating 16-bit sequence, an oscilloscope trace of which is shown in figures \ref{fig:Encoder}b) and c). Finally, the two paths merge again at the Wollaston polarizer and exit through the circulator, with the output polarization determined by the applied relative phase $\varphi$.\\
The Wollaston polarizer is a key element for achieving the reported low \acp{QBER}: it offers a nominal extinction ratio of $10^6$, orders of magnitude larger than comparable commercially available fiber-based components. Additionally, it acts not only as an initial polarizing beamsplitter, but also as a filter of equal quality upon merging the pulses again, as photons with an incorrect polarization are refracted away from the intended beam path and not sent back to the circulator. This double functionality makes the setup robust to polarization misalignment within the loop: deviations from the optimal polarization states always affect both propagation directions equally and therefore do not result in higher error rates, but only in reduced transmission efficiency.\\
The aforementioned features combine to a setup that not only yields the excellent performance reported on here, but is comparably easy to use. No special care is required in its vicinity with regard to mechanical or temperature control (see section \ref{subsec:StabilityResults}), and polarization alignment within the loop consists simply of rotating the waveplates in each arm iteratively to maximize the output of the setup, eliminating the need for tedious extinction-ratio or polarization state measurements.

\subsection{Decoding and Characterization} \label{subsec:Decoding}
\begin{figure*}[hbtp]
    \centering
    \includegraphics[width=\linewidth]{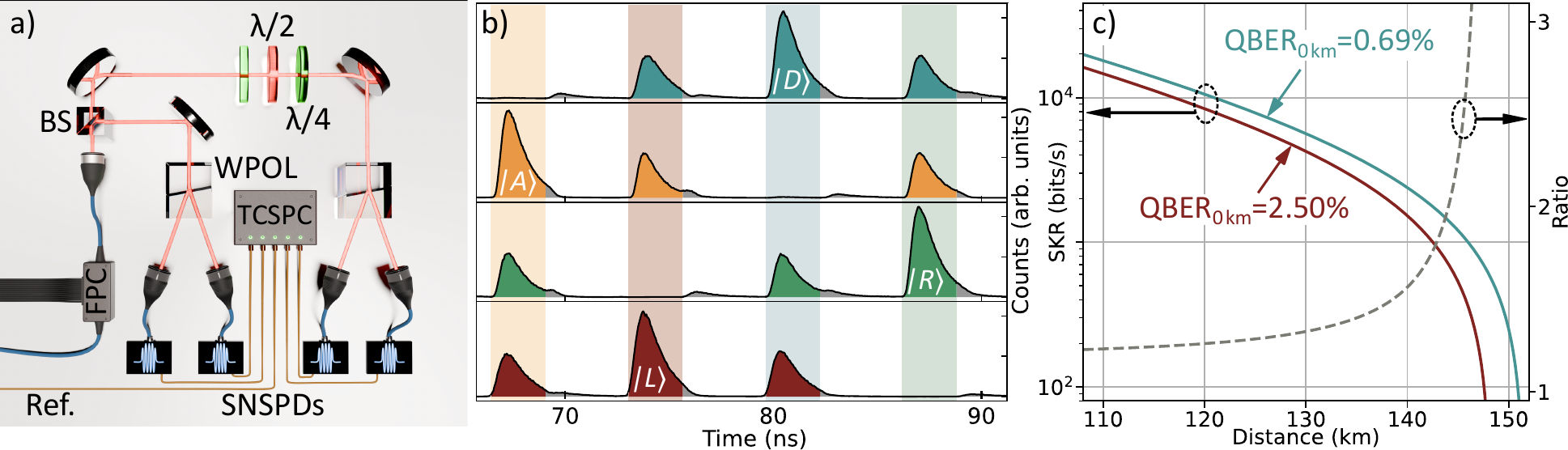} 
    \caption{\textbf{Low-error polarization encoding and decoding of single photons from the \ac{QD}.} a) Optical setup for polarization decoding of BB84 states. FPC: fiber polarization controller; BS: 50/50 beamsplitter; WPOL: Wollaston polarizer; TCSPC: time-correlated single photon counter; Ref.: reference signal from \ac{AWG}. b) Histograms showing event detection times for all channels across four bits of the repeating sequence relative to the reference signal. Colored parts show counts used for \ac{QBER}/encoding agreement calculations, grey parts are rejected. Temporal filtering is necessary here as the time needed for a complete decay of the \ac{QD} signal exceeds the $3.3\,\unit{ns}$ limit. c) Influence of low \ac{QBER} on the achievable asymptotic \ac{SKR} over distance in a simulated BB84 experiment (see Methods section for details). The gray, dashed line shows the relative increase in \ac{SKR} between our error rate of $0.69\%$ and a generic error rate of $2.50\%$ for otherwise equal experimental parameters.}
    \label{fig:Decoder}
\end{figure*}

To quantify the performance of the \ac{FSSI} encoder, two different methods are employed. First, to demonstrate the successful encoding of single photons from the \ac{QD}, the encoded photons are sent to a home-built BB84 polarization decoding setup (see Fig. \ref{fig:Decoder}a)). Second, in order to confirm the expected stability of the output polarization, a long-term measurement is conducted using a continuous-wave laser and a polarimeter (Thorlabs PAX1000IR2).\\
The polarization decoder is a refined version of a setup used in a previous experiment \cite{yangHighrateIntercityQuantum2024}, and -- like the encoder -- uses high quality free space polarization optics to ensure low loss and accurate decoding. The arriving photons first pass through an electronic, fiber-based 4-channel polarization controller (GeneralPhotonics PCD-M02) to compensate for polarization changes caused by the connection fiber. The random basis choice is made passively by a 50/50 free space beamsplitter. In each basis, a Wollaston polarizer is used to split polarization states. The four outputs are sent to four channels of the \ac{SNSPD}, and detected events are temporally correlated with a timing reference from the \ac{AWG} at the beginning of each sequence.
\begin{figure}[H]
    \centering
    \includegraphics[width=\linewidth]{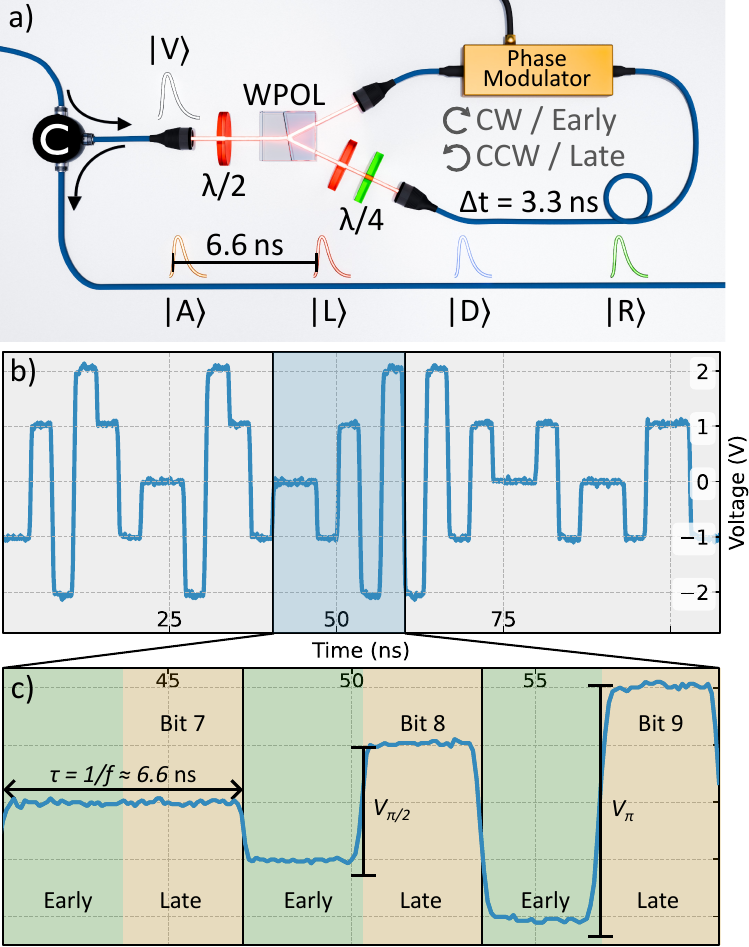}
    \caption{\textbf{Polarization encoding method.} a) Polarization encoding setup. Photons coming from the extraction setup are coupled to a polarization-maintaining circulator, which directs them to the Sagnac loop and, after propagation through the loop, to the output fiber. The \ac{AWG} that drives the \ac{PM} also drives the excitation laser, ensuring continuous synchronization between excitation and encoding. WPOL: Wollaston polarizer; C: circulator; (C)CW: (counter)clockwise. b) Voltage signal applied to the phase modulator to encode a repeating 16-bit polarization sequence, as measured on an oscilloscope. c) Zoom-in view on the shaded area in b). Each time slot is split into an early and a late part, corresponding to the presence of \ac{CW} and \ac{CCW} photons within the \ac{PM}.}
    \label{fig:Encoder}
\end{figure}
During test measurements in which the single photons passed through the encoding and decoding setups (with the phase modulator switched off), a static-encoding \ac{QBER} of $0.11(1)\%$ is obtained after optimizing the fiber polarization controller to the $\ket{D}$ state. This value is limited by the dark counts of the \acp{SNSPD}, which yield $\sim 160\text{cts/s}$ with fibers connected but the \ac{QD} signal blocked.\\
For the stability measurement, an additional free space 50/50 beamsplitter (not shown in Fig. \ref{fig:Encoder}) was inserted into the encoder between the input/output coupler and the $\lambda/2$ waveplate. Two additional mirrors were then used to align the output light from the Sagnac loop to hit the polarimeter sensor precisely in the center and under normal incidence. This configuration was chosen since it allowed avoiding all fibers outside of the Sagnac loop during the measurement, such that the actual encoding stability itself could be measured isolated from fiber-induced polarization fluctuations. In practical scenarios this isolated encoding stability is the quantity of interest: Variations induced by fibers after the encoder can be easily corrected, but those originating from the encoder itself (for instance due to uneven splitting between \ac{CW} and \ac{CCW} components) cannot.

\section{Performance} \label{sec:Results}
\subsection{Encoding and Decoding of Single Photons} \label{subsec:EncodingResults}
\begin{figure*}[htbp]
    \centering
    \includegraphics[width=\linewidth]{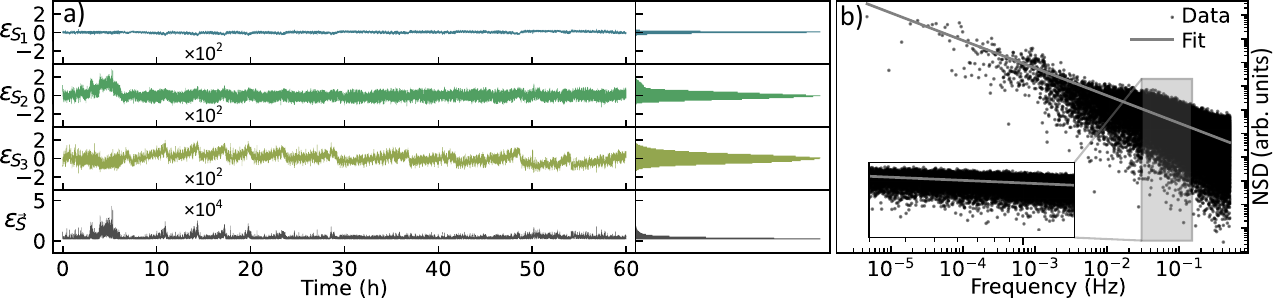}
    \caption{\textbf{Polarization stability measurements.} a) Deviations in normalized Stokes components over time, defined as $\varepsilon_{S_i}(t)=S_i(t)-\overline{S}_i$. Bottom: Deviations in the projection of the Stokes vector at time $t$ onto the average Stokes vector. b) Polarization noise spectral density (NSD) of $\varepsilon_{\vec{S}}$ (double-logarithmic scale). A fit according to $c\cdot f^{-\beta}$ yields $\beta=1.16$, showing that the noise spectrum is dominated by flicker noise (where $\beta=1$). We attribute this noise to the photodiode inside the polarimeter \cite{weissman1fNoiseOther1988}. Inset: zoom-in on shaded area, showing good agreement between fit and data especially for higher frequencies. All data was recorded for an inactive phase modulator ($\ket{D}$ state), and datapoints were acquired in $1\,\unit{s}$ intervals at a power of $200\,\text{µW}$ as measured by the polarimeter.}
    \label{fig:StabilityResults}
\end{figure*}
In order to demonstrate the encoding of single photons from the \ac{QD}, the performance of the system for encoding a pseudorandom 16-bit repeating polarization sequence (L,A,R,D,A,R,D,L,A,A,L,D,R,D,L,R) is examined. A repeating sequence was chosen to show that the system is capable of modulating random sequences with high reliability for all four states simultaneously.\\ 
In Fig. \ref{fig:Decoder}b), the resulting temporal correlations between the respective decoder output channels and an electronic reference signal from the \ac{AWG} (sent at the beginning of each sequence) are shown. They demonstrate that the photon polarization is encoded as desired: for instance, each $\ket{D}$-encoded photon is decoded and detected as $\ket{D}$ with a near $50\%$ probability, as $\ket{A}$ with a near $0\%$ probability, and as $\ket{R}$ or $\ket{L}$ with a near $25\%$ probability each, as required by the BB84 protocol \cite{bennettQuantumCryptographyPublic2014}. Calculating the \ac{QBER} for the entire 16-bit sequence yields a mean value of $0.69(2)\%$ ($0.96(1)\%$ for the $X$-basis, $0.43(2)\%$ for the $Y$-basis). Upon reconstructing the experimental encoded state matrix from the measured data and comparing it to the theoretically expected one, one obtains an encoding agreement of $96.5(2)\%$. 
Calculations are described in more detail in the Methods section.\\
Here, data is accumulated for only one minute in order to minimize the effects of polarization fluctuations in the connection fiber. Temporal filtering is used to avoid effects of late \ac{CCW} photons being modulated as \ac{CW} ones and vice versa, such that a high repetition rate can be maintained. While temporal filtering discards roughly $10\%$ of events, this quantity is not factored into the transmission efficiency, as it is not inherent to the encoding module but rather a limitation of the used \ac{SPS} and repetition rate. Nonetheless, we acknowledge that single-pass or \ac{MZI} encoders would support higher repetition rates than a Sagnac-based one before being met with this limitation.\\
Transmission efficiencies are $\eta_{\text{dec}}=81.3\%$ for the decoder (including the connection fiber with which the FPC is integrated) and $\eta_{\text{enc}}=\text{30.4}\%$ for the encoder, with main sources of loss being the insertion losses of the \ac{PM} and circulator. An overview of relevant parameters is provided in table \ref{tab:Params}.\\

\begin{table}
    \caption{Relevant parameters of involved optical components and setups. Note that individual losses add up to more than the total setup loss of the encoder; we attribute these additional losses to fiber connections needed for the insertion loss measurements. Further note that fiber coupling losses were estimated by using the setup to couple to fibers of the same type as the ones used for the \ac{PM}/circulator, in order to isolate coupling losses from insertion losses. All values were measured with a \ac{CW} laser at the \ac{QD} emission wavelength.}
    \begin{tabular}{ccc}
        \hline
        Parameter & Description & Value \\
        \hline
        $\lambda_{\text{exc}}$ & Excitation laser wavelength & $1533.8(1)\,\unit{nm}$\\
        $\lambda_{\text{QD}}$ & \ac{QD} emission wavelength & $1560.4(1)\,\unit{nm}$\\
        $f_{\text{exc}}$ & System repetition rate & $151.894\,\unit{MHz}$\\
        $L_{\text{circ},12}$ & Circulator insertion loss (port $1\rightarrow2$) & $1.21(5)\,\unit{dB}$\\
        $L_{\text{circ},23}$ & Circulator insertion loss (port $2\rightarrow3$) & $0.86(4)\,\unit{dB}$\\
        $L_{\text{PM}}$ & \ac{PM} insertion loss & $2.29(5)\,\unit{dB}$\\
        $L_{\text{C}}$ & Estimated total fiber coupling loss & $1.3(1)\,\unit{dB}$\\
        $L_{\text{tot}}$ & Total encoder loss & $5.17(5)\,\unit{dB}$\\
        $\text{ER}_{\text{\ac{PM}}}$ & \ac{PM} extinction ratio & $27.7(3)\,\unit{dB}$\\
        \hline
    \end{tabular}
    \label{tab:Params}
\end{table}
In order to highlight the application potential of the module, Fig. \ref{fig:Decoder}c) shows an exemplary simulation for the asymptotic \ac{SKR}\cite{gottesmanSecurityQuantumKey2004, yangHighrateIntercityQuantum2024} in a hypothetical BB84 \ac{QKD} experiment. For otherwise equal parameters, a zero-distance \ac{QBER} of $0.69\%$ yields a $>19\%$ increase in \ac{SKR} for all distances and an increase of $0.63\,\unit{dB}$ in maximum tolerable loss. The compared \ac{QBER} of $2.50\%$ was chosen as a lower bound of values reported for high-speed \ac{SPS} encoding experiments so far; for details, see table \ref{tab:Comparison}.\\
At this point it should be mentioned that in the setting described here, temporal filtering may induce a potential security risk. Very recently, a study revealed the threat of temporal side-channel information leakage in Sagnac-based \ac{QKD} encoders \cite{tanExperimentalStudyTimedependent2025}. While this study examines weak coherent sources as opposed to sub-Poissonian ones and its results can therefore not be directly transferred to our work, we acknowledge the potential presence of time-dependent side channel vulnerabilities in our system and encourage further research on this topic.

\begin{table*}[]
    \centering
    \caption{Comparison of our setup to other reported, comparable \ac{SPS}-encoding experiments. Reference \cite{collinsQuantumKeyDistribution2010} is included for the sake of completeness, although they report $g^{(2)}(0)>0.5$. For this reason, the reported value is not considered in Fig. \ref{fig:Decoder}b).}
    \begin{tabular}{c c c c c c c c c}
        \hline
        Reference & Source & Protocol & Type & Method & Rate ($\unit{MHz}$) & \ac{QBER} ($\%$) & Ch. loss\\
        \hline
        This work & \ac{QD} (C-band) & BB84 & Polarization & Sagnac & $151.894$ & $0.69$ & --\\
        \cite{waksQuantumCryptographyPhoton2002} & \ac{QD} & BB84 & Polarization & Single-pass & $76$ & $2.5$ & --\\
        \cite{intalluraQuantumKeyDistribution2007, intalluraQuantumCommunicationUsing2009} & \ac{QD} (O-band) & BB84 & Time-bin & \ac{MZI} & $1$ & $5.9$ & $12.8\,\unit{dB}$\\
        \cite{collinsQuantumKeyDistribution2010} & \ac{QD} ($894\,\unit{nm}$) & BB84 & Polarization & Single-pass & $40$ & $1.2$ & --\\
        \cite{heindelQuantumKeyDistribution2012} & \ac{QD} ($897\,\unit{nm}$) &  BB84 & Polarization & Single-pass & $200$ & $3.8$ & --\\
        \cite{takemotoQuantumKeyDistribution2015} & \ac{QD} (L-band) & BB84 & Time-bin & \ac{MZI} & $62.5$ & $2.3$ (est.) & --\\
        \cite{samanerFreeSpaceQuantumKey2022} & hBN defects & B92 & Polarization & Single-pass & $1$ & $8.95$ & --\\
        \cite{al-jubooriQuantumKeyDistribution2023} & hBN defects & BB84 & Polarization & Single-pass & $0.5$ & $\geq3$ & --\\
        \cite{zahidyQuantumKeyDistribution2024} & \ac{QD} ($924\,\unit{nm}$) & BB84 & Polarization & Single-pass & $72.6$ & $\geq3.25$ & $9.6\,\unit{dB}$\\
        \cite{zhangExperimentalSinglePhotonQuantum2025c} & \ac{QD} ($885\,\unit{nm}$) & BB84 & Polarization & Single-pass & $76.13$ & $2.54$ & --\\
        \cite{vajnerSinglePhotonAdvantageQuantum2024} & \ac{QD} ($921\,\unit{nm}$) & \ac{QSCF} & Polarization & Single-pass & $80$ & $2.8$ & --\\
    \end{tabular}
    \label{tab:Comparison}
\end{table*}

\subsection{Stability} \label{subsec:StabilityResults}
In Fig. \ref{fig:StabilityResults}, the temporal evolution of the output polarization of the encoder module is shown. This measurement was conducted over an extended timespan of $60$ hours and highlights the stability of the module without the need for active temperature control or mechanical isolation. In Fig. \ref{fig:StabilityResults}a), the quantity
\begin{equation}
\varepsilon_{\vec{S}}(t)=1-\vec{S}(t)\cdot\vec{S}_{\text{avg}}
\end{equation}
is shown in the bottom plot and used as a measure of error in the output polarization of the setup for a fixed input polarization. This yields an average polarization error of only $4\times10^{-5}$ and a maximum polarization error of $4.3\times10^{-4}$ over the entire measurement duration. In Fig. \ref{fig:StabilityResults}b), a discrete Fourier transform on this quantity reveals that the noise present in the signal is dominated by flicker noise proportional to $1/f$, and that no particularly strong noise frequency is present in the signal even for low frequencies.\\
The measurements were conducted in a temperature-regulated laboratory and with the module placed on a vibration-isolated optical table. However, along the $60$ hour measurement duration, regular laboratory works were performed in the same room and on the same optical table, and even the input fiber of the encoder was moved; creating disturbances much stronger than e.g. an \ac{MZI} could typically tolerate. While changes in the polarization state are visible (especially within the first $\sim6$ hours of measurement), they are negligible compared to other sources of error in our setup such as imperfections in the electronic pulses or human inaccuracy during manual optimization. Additionally, they appear to be reversible in time, as all measured parameters continuously tend back towards their initial settings.

\section{Conclusion} \label{sec:Conclusion}
We have presented a high-quality encoding module for single photons at telecom wavelength, based on a novel free-space Sagnac-loop configuration. This is the first demonstration of such a free-space system capable of encoding polarization information on sub-Poissonian photons from a deterministic \ac{SPS}. The system achieves a high encoding agreement of $96.5(2)\%$ and a low \ac{QBER} of $0.69(2)\%$, representing the best results reported at present regarding dynamic information inscription onto single photons. Within the experiment, these values are obtained at a high repetition rate of $152\,\unit{MHz}$, and could be improved further by implementing dynamic feedback by monitoring those quantities in real time. We have also demonstrated the long-term stability of the output states over a time span of $2.5$ days of continuous operation, without the need for active mechanical or thermal stabilization. This enables operational simplicity and reduces calibration process overhead for practical applications. Despite the encoding module exhibiting a loss of $5.17(5)\,\unit{dB}$, a reduction of $\sim2\,\unit{dB}$ can be reasonablely anticipated by reducing fiber coupling losses through the deployment of a free space circulator. This approach facilitates direct free-space coupling into the fiber-based \ac{PM} and the integration of single-photon collection setup with the encoding module in free space. \\
Beyond polarization encoding depicted in this work, we envision three additional potential applications of this encoding protocol as a summary. First, all components being used have integrated photonics counterparts \cite{suFourportIntegratedPolarizing2014, sibsonIntegratedSiliconPhotonics2017, avesaniFullDaylightQuantumkeydistribution2021}, rendering the setup feasible for future on-chip integration. Second, with minor modifications, the current architecture could be adapted for phase and time-bin encoding  \cite{wangPracticalGigahertzQuantum2018} while retaining aforementioned performance. Third, the system can operate as a active single-photon decoder of equivalent performance by simply reversing the input and output ports, enabling for instance \ac{QKD} protocols with adaptable, asymmetric basis choice \cite{loEfficientQuantumKey2005a}. The widespread applicability of this design featuring high precision, long-term stability and operational simplicity, marks a significant advancement in the field of single-photon information processing and positions it as a strong candidate for future standardization in quantum network applications.

\section*{Acknowledgements}
We thank Jialiang Wang for fruitful discussions on matters of electronics, Fabian Klingmann for assisting in the initial design and assembly of the decoder module, Vincent Rehlinger for helpful insights about the requirements of practical \ac{QKD} implementations, and Dan Huy Chau for refining the optical setup renderings. We thank the companies Swabian Instruments, PriTel, EXFO, and Active Technologies for their continued and timely support. The authors gratefully acknowledge the funding by the German Federal Ministry of Education and Research (BMBF) within the project QR.N (16KIS2188), SQuaD (16KISQ117) and SemIQON (13N16291), the European Research Council (MiNet GA101043851), the EMPIR programme co-financed by the participating states, the Deutsche Forschungsgemeinschaft (DFG, German Research Foundation) within the project InterSync (GZ: INST 187/880-1 AOBJ: 683478), and under Germany’s Excellence Strategy (EXC-2123) Quantum Frontiers (390837967), and Flexible Funds programme by Leibniz University Hannover (51410122).

\section*{Author Contributions}
J.H. constructed the encoder and decoder modules and conducted and analyzed the presented measurements. Z.J., F.B. and J.Y., with support from R.J., constructed the extraction setup and performed the \ac{QD} excitation. J.W. supported the experiment in matters of electronics, and the conceptualization of the idea for the encoder setup. T.F. developed software for analyzing the measurement data. E.P.R. provided support with instrumentation and optical experiments. M.J., S.L.P., and P.M. designed and fabricated the \ac{QD} source. J.H. prepared the first manuscript draft and revised it with the help of J.Y., E.P.R., S.L.P., M.Z., P.M., F.D. and all other co-authors. J.Y., M.Z., and F.D. conceived the idea, supervised the project and acquired relevant funding.

\section*{Conflict of Interest}
The authors declare no conflicts of interest.

\appendix
\section*{Methods}
\subsection*{Second Order Autocorrelation}
\label{app:SecondOrderCorrelation}
For computing the value of $g^{(2)}(0)$ presented above, the recorded second-order autocorrelation data was fitted with a function consisting of side-peak terms $f_{S,i}$, a central-peak term $f_C$, and a bunching term $f_B$. Each side-peak term consists of a convolution of a Gaussian to model the \ac{IRF} and a two-sided exponential decay,
\begin{equation}
    f_{S,i}(t)=A_S\left(\exp(-t^2/(2\sigma_t^2)) * \exp(-\left| t-t_0+i/f_\text{Sys}\right|) / T_D \right),
\end{equation}
where $A_S$ is the pulse amplitude, $\sigma_t$ is the \ac{IRF}, $t_0$ accounts for potential offset of measurement data in the $t$ direction, $f_{\text{Sys}}=151.894\,\unit{MHz}$ is the system repetition rate, $i\neq0$ is the position of the side peak, and $T_D$ represents the decay time of the \ac{QD} emission. The term $f_C$ is largely identical, but has $i=0$ and an independent amplitude $A_C$ instead of $A_S$. Finally, the bunching term is of the form
\begin{equation}
    f_B=1+A_b\exp(-\left|t-t_0\right|/T_B),
\end{equation}
where $T_B$ is the bunching time (we find $T_B=(38\pm12)\,\unit{ns}$). The full fit function is then of the form
\begin{equation}
    g^{(2)}(t)=\left(f_C(t)+\sum\limits_if_{S,i}\right)\cdot f_B,
\end{equation}
and the value for $g^{(2)}(0)$ is calculated as
\begin{equation}
    g^{(2)}(0)=\nicefrac{\int f_C(t)\,\text{d}t}{\int f_S(t)\,\text{d}t},
\end{equation}
with both terms centered at $t=0$ and integration limits of $\pm100\,\unit{ns}$.

\subsection*{QBER and Encoding Agreement}
\label{app:QBERandFidelity}
The \ac{QBER} and encoding agreement are calculated as follows. First, integrating the counts in each channel over a $2.65\,\unit{ns}$ temporal filtering window within each time slot $i$ yields a total of $64$ values $n_{\ket{\psi},i}$, where $\ket{\psi}\in\{\ket{D},\ket{A},\ket{R},\ket{L}\}$ represents the possible polarization states and $i=1,2,\ldots16$ represents the respective bit in the sequence. The \ac{QBER} is computed for each time slot individually by only considering counts $c_{\ket{\psi},i}$ in the expected measurement basis for that slot, for instance
\begin{equation}
    \text{QBER}_7=c_{\ket{A},7}/(c_{\ket{D},7}+c_{\ket{A},7})
\end{equation}
for bit $7$ in which the $\ket{D}$ state is encoded. When averaging over all bits in the sequence, this yields the presented mean \ac{QBER} of $0.69(2)\%$. A value of $0.96(1)\%$ is obtained when only averaging over events where the $X$-basis was encoded, $0.43(2)\%$ for the $Y$-basis.\\
For the encoding agreement, the same $64$ values obtained after temporal filtering and integration serve as a starting point. First, four experimental four-dimensional state vectors $\ket{\psi}_{\text{exp}}$, one for each encoded state, are constructed from the data by adding up integrated measured counts from each channel for events where that specific state was encoded. For instance, the experimental vector $\ket{D}_{\text{exp}}$ is computed as
\begin{equation}
    \ket{D}_{\text{exp}}=\sum\limits_{i=4,7,12,14}(c_{\ket{D},i},\,c_{\ket{A},i},\,c_{\ket{R},i},\,c_{\ket{L},i}),
\end{equation}
as the $\ket{D}$ state is encoded in bits 4, 7, 12, and 14 of the sequence. 
These state vectors are then normalized and the experimental encoding matrix $M_{\text{exp}}$ is constructed from them simply by using each state vector as one column.\\
The theoretical encoding matrix $M_{\text{theo}}$ is obtained likewise from the normalized theoretically expected state vectors $\ket{\psi}_{\text{theo}}$ (in the corresponding order), where \begin{equation}
    \ket{D}_{\text{theo}}=\sqrt{3/2}\cdot(1,\,0,\,1/2,\,1/2)
\end{equation}
and so on. Finally, the encoding agreement is computed as
\begin{equation}
    F=\left. 1-\sqrt{\sum\limits_{i=1}^{4}\sum\limits_{j=1}^{4}(M_{\text{exp}}^{i,j}-M_{\text{theo}}^{i,j})^2}\middle/\sqrt{\sum\limits_{i=1}^{4}\sum\limits_{j=1}^{4}(M_{\text{theo}}^{i,j})^2}. \right.
\end{equation}
The subtracted term describes the distance between the experimental and theoretical matrices in terms of the Frobenius norm, relative to that same norm applied to the theoretical matrix.

\subsection*{\ac{SKR} Simulations}
\label{app:SKR}
The asymptotic \ac{SKR} $A$ is calculated the same way as in \cite{yangHighrateIntercityQuantum2024}, namely as
\begin{equation}
    A = R \cdot p_{\text{sift}} \cdot \lbrace \underline{p}_c^{(1)}\cdot\left(1-h(\overline{e}_1)\right)-f_\text{EC}\cdot p_c \cdot h(e_\text{tot}) \rbrace,
\end{equation}
where $R=151.894\,\unit{MHz}$ is the system repetition rate, $p_{\text{sift}}=0.5$, $\underline{p}_c^{(1)}$ is the lower bound of single photon event detection probability per time slot, $\overline{e}_1$ is the upper \ac{QBER} bound for single-photon states, $f_{\text{EC}}=1.16$ is the error correction efficiency, $p_c$ is the total detection probability per time slot across all photon number states, $e_{\text{tot}}$ is the total \ac{QBER} across all photon number states, and $h(\cdot)$ is the binary entropy function.\\
Assuming that $n$-photon events are negligible for $n>2$, we get
\begin{equation}
    p_2=\dfrac{\langle n \rangle ^2 \cdot g^{(2)}(0)}{2},\quad p_1=\langle n \rangle - p_2,\quad p_0=1-p_1-p_2.
\end{equation}
The detector click probability for an $n$-photon state (neglecting dead times) is then given as
\begin{equation}
    p_c^{(n)}=p_n\left[1-(1-p_{\text{dc}})\cdot(1-\eta_E\eta_C\eta_D\eta_S)^n\right],
\end{equation}
where $p_{\text{dc}}=50/R$ is the probability to detect a dark count in a given time slot, and $\eta_{E(C,D,S)}$ is the transmission efficiency of the encoder module (fiber channel, decoder module, single photon detector). Then,
\begin{align}
    e_n&=\dfrac{p_{\text{dc}}/2+p_{\text{mis}}\cdot(1-(1-\eta_E\eta_C\eta_D\eta_S)^n)}{1-(1-p_{\text{dc}}\cdot(1-\eta_E\eta_C\eta_D\eta_S)^n},\\
    e_{\text{tot}}&=\dfrac{1}{p_c}\sum_{n=0}^{2}e_np_n\left(1-(1-p_{\text{dc}})\cdot(1-\eta_E\eta_C\eta_D\eta_s)^n\right),\\
    \overline{e}_1&\leq\dfrac{e_{\text{tot}}p_c-p_0p_{\text{dc}}/2}{1-(1-p_{\text{dc}})\cdot(1-\eta_E\eta_C\eta_D\eta_S)},\quad\text{and}\\
    \underline{p}_c^{(1)}&\geq p_c-p_0p_{\text{dc}}-p_2.
\end{align}
In Fig. \ref{fig:Decoder}, the quantity $p_{\text{mis}}$ is varied between $0.69\%$ and $2.50\%$. Although this quantity is not strictly equal to the \ac{QBER}, the \ac{QBER} provides an upper bound and constitutes a good approximation to $p_{\text{mis}}$ as long as the ratio between signal counts and detector dark counts is sufficiently large. The remaining parameters in the simulations are given in table \ref{tab:SKR_params}. Note that the detector parameters in the simulation are better than those in the experiment, to more closely represent the state of the art in \acp{SNSPD} and enable a more accurate comparison.

\begin{table}[]
    \centering
    \caption{Parameters for asymptotic \ac{SKR} simulations.}
    \begin{tabular}{c c c c}
        \hline
        Parameter & Value & Parameter & Value \\
        \hline 
        $\eta_E$ & 0.34 & $\eta_C$ & $0.18\,\unit{dB}/\unit{km}$\\
        $\eta_D$ & 0.8 & $\eta_S$ & 0.8\\
        $\langle n \rangle$ & $0.138$ & $g^{(2)}(0)$ & $0.005$\\       
        \hline
    \end{tabular}
    \label{tab:SKR_params}
\end{table}


\end{document}